**Quantitative Detection of Molecular Oxygen in the Gas Phase with Fluorescent Nanodiamonds**

Nicholas A. Nunn*[1], Antonin Marek[1], Alex I. Smirnov[2], Olga A. Shenderova[1], Marco D. Torelli*[1]

[1]Adámas Nanotechnologies, Inc., Raleigh, NC, USA

[2]Department of Chemistry, North Carolina State University, Raleigh, NC, USA

*Corresponding Authors:

nnunn@adamasnano.com (ORCID: 0000-0002-2021-2575)

mtorelli@adamasnano.com

**Abstract**

The quantitative detection of paramagnetic molecular oxygen ($O_2$) in gas mixtures using optically detected magnetic resonance (ODMR) from negatively charged nitrogen-vacancy ($NV^-$) centers in fluorescent nanodiamonds is described. Fluorescent nanodiamonds approximately 70 nm in diameter were deposited on the glass surface of a microfluidic channel, and the oxygen concentration varied from 0 to 100% (0 to 760 mmHg $O_2$ partial pressure) by mixing $O_2$ and $N_2$ gases at ambient pressure. Continuous-wave (CW) ODMR contrast was measured using a double-modulation (lock-in) detection scheme applied to both optical excitation and microwave drives. The ODMR contrast decreases linearly with oxygen partial pressure, with a sensitivity coefficient $k$ of (-10.1×± 0.3) $\times 10^{-4}$ % $mmHg^{-1}$. The oxygen detection limit of the experimental setup was estimated to be approximately 8 mmHg $O_2$ partial pressure (corresponding to about 1% $O_2$ in the gas mixture). Cycling of the content of $O_2$ in the gas mixture in the range of 0-5 % revealed a slight hysteresis and corresponding repeatability of 0.006 in percent ODMR contrast. The observed fluorescence quenching and relatively slow response (ranging from several to tens of minutes) upon changes in oxygen concentration suggest that physisorption of gas molecules on the nanodiamond surfaces contributes to equilibration dynamics. The applicability of the nanodiamond-based oxygen quantum sensor was further demonstrated by detecting transient bursts of molecular oxygen generated by enzyme-catalyzed decomposition of hydrogen peroxide.

## 1. Introduction

As a quantum sensing material, fluorescent nanodiamond containing nitrogen-vacancy ($NV^-$) centers has the capability to improve sensitivity, resolution, and speed compared to classical-based measurements.[1-3] $NV^-$ centers possess uniquely coupled magneto-optical properties,[4] where quantum control and optical readout of the electronic spin state have been successfully employed to measure magnetic fields, temperature, physical orientation, and paramagnetic chemical analytes.[4-7] The exceptional photostability of $NV^-$ centers[8] in combination with the biocompatibility[9, 10] of diamond make it a powerful optical sensor platform. While both macroscopic single-crystal diamond plates and nanodiamond particles have been actively explored for quantum sensing applications, nanoparticles provide a more cost-effective platform[11] and, owing to their nanoscale dimensions, can be readily deployed for intracellular measurements.[7]

Paramagnetic species such as transition metal ions and free radicals are key players in myriads of chemical and biochemical processes; therefore, methods that facilitate their detection, including those based on $NV^-$ quantum sensing, are highly attractive. While gadolinium ion ($Gd^{3+}$)–based MRI contrast agents are among the most extensively studied paramagnetic species because this ion possesses $S$=7/2 electronic spin state arising from seven unpaired 4*f* electrons,[12-16] an expanding body of research has focused on other transition metals such as manganese[17, 18] and iron,[18-21] which play a major role in oxygen transport and metabolism.[22-24] Further development of $NV^-$ based methods to detect paramagnetic analytes has extended to the measurement of oxygen-based free radicals, whose dysregulation contributes to progression of disease through an onset of oxidative stress[25-27] (*e.g.*, cardiovascular diseases,[28] Alzheimer's disease,[29, 30] cancer,[31] *etc.*). Recently, measurements have extended beyond *in situ* proof-of-principle demonstrations of stable free radicals[32, 33] and ROS models[34-38] into disease models including inflammation,[39, 40] atherosclerosis,[41] and Huntington's disease.[42] This unique capacity to spatially resolve cellular redox mediators can help to establish relationships in disease pathogenesis through a detailed mapping of paramagnetic species location and concentrations.[43, 44]

While efforts to develop $NV^-$-based quantum sensors of paramagnetic metal ions and ROS continue to grow, measurement of widely present paramagnetic molecular oxygen has been under-reported. Oxygen is a stable paramagnetic diradical, which has two unpaired electrons in degenerate $\pi^*$ orbitals in its triplet ground state.[45] Oxygen plays a central role in aerobic metabolism as the terminal electron acceptor in the mitochondrial respiratory chain. In the body, oxygen is transported by hemoglobin-bound iron in red blood cells and it serves as a critical precursor to ROS through partial reduction pathways.[46] One of the most powerful approaches to measure oxygen in biological samples and tissues is based on Electron Paramagnetic Resonance (EPR) spectroscopy.[47] The EPR oximetry method is based on magnetic interactions between oxygen and a paramagnetic spin probe (typically a stable radical) that results in shortening of electronic $T_1$ and $T_2$ relaxation times of the probe. These interactions are thought to be dominated by Heisenberg spin exchange instead of dipolar interactions.[48] The electronic spin of oxygen ($S$=1) is also a source of spin-noise, which is expected to affect quantum sensing of other paramagnetic species in biological systems. In the gas phase, exposure of diamond surfaces to air improves $NV^-$ optical properties compared to vacuum, either through $O_2$ adsorption or oxygen-containing dielectric surface passivation.[49] Yet relaxation times of sub-surface $NV^-$ centers in bulk single crystal diamond are shortened in the presence of dissolved oxygen.[50, 51] However, recent works from the Schirhagl and Simpson groups on $NV^-$-containing nanodiamond particles reported that $T_1$ relaxation was not affected by dissolved oxygen.[52,53] To the best of our knowledge, there are no prior studies demonstrating the capability of $NV^-$ centers to detect molecular $O_2$ in the gas phase.

In this work, we examine the effect of gaseous oxygen on $NV^-$ centers in nanodiamond particles by continuous wave optically detected magnetic resonance (CW-ODMR). A double-modulation (lock-in) detection scheme applied to both optical excitation and microwave drives was employed to increase sensitivity and allowed for real-time readout. We demonstrate ambient temperature measurements of molecular oxygen titrated with nitrogen over a wide dynamic range from 0 to 760 mmHg partial pressure of $O_2$ and achieve a detection down to 1% $O_2$ v/v.

## 2. Experimental

### 2.1 Materials

Fluorescent diamond particles containing an estimated 2 ppm of $NV^-$ centers (NDNV70nmHi, Adámas Nanotechnologies, Inc., Raleigh, NC, USA) were used for experiments. Ultra-High Purity grade (99.999%) Nitrogen ($N_2$) and Industrial Grade Air were sourced from Airgas (Airgas USA LLC, Radnor, PA, USA). High purity (≥99.5%) Oxygen ($O_2$) was sourced from Linde (Linde Gas & Equipment Inc., Burr Ridge, IL, USA). Lab grade 30% (w/w) hydrogen peroxide was sourced from Lab Alley Chemicals (Lab Alley Essential Chemicals, Austin, TX, USA) and Catalase (*A. niger*) (HY-135849) was sourced from MedChemExpress (MedChemExpress LLC., Monmouth Junction, NJ, USA). A glass bottom Luer flow cell μ-Slide with a flow channel of dimensions 0.25 mm (height) × 5 mm (width) × 50 mm (length) (μ-Slide $I^{0.2}$ Luer Glass Bottom - 80167) was sourced from Ibidi (Ibidi GmbH, Gräfelfing, Germany). Diamond particles were deposited onto the glass surface of the slide by injecting 100 μL of a 30 mg/mL aqueous suspension of the diamond particles into the channel. The slide was heated on a hotplate at 70 °C overnight, leaving a dry film of diamond particles inside the channel of the slide.

### 2.2 Optical and Microwave Instrumentation

Fluorescence signals were recorded with the Adamas ODMR-Pulse system (Adámas Nanotechnologies, Inc., Raleigh, NC, USA). All hardware components were controlled with proprietary software coded in MATLAB® (The MathWorks, Inc., Natick, MA, USA). Light excitation was provided by an 8.6 W LED (ThorLabs, Inc., Newton, NJ, USA) with a dominant excitation wavelength of 565 nm. The LED was operated at its maximum output for all measurements. The excitation light passes through a filter cube with a 562 nm band pass excitation filter (FF01-562/40, Semrock, Inc., Rochester, NY, USA), 593 nm dichroic beamsplitter (FF593-Di03, Semrock, Inc.), and a 650 nm longpass emission filter (FELH0650, ThorLabs, Inc.). The 650 nm longpass emission is to ensure that the detected fluorescence is primarily from negatively charged NV centers. Optical excitation and detection elements were integrated with an Olympus IX-71 inverted epifluorescence microscope (Olympus Corporation of the Americas, Center Valley, PA, USA). Microwaves were produced by a signal generator (Standard Research Systems, Sunnyvale, CA, USA) and swept or modulated by control via an arbitrary wave generator (Siglent

Technologies Co., Ltd., Shenzhen, CN). Microwaves were delivered to the sample by a previously reported PCB microwave antenna with a 1 mm aperture affixed to a 3D printed microscope stage insert provides microwaves at the sample.[54]

## 2.3 Oxygen Detection Experiments

### *2.3.1 Gas Mixtures*

Two experiments are reported: (i) controlled gas mixtures of either nitrogen ($N_2$) and pure oxygen ($O_2$) or $N_2$ and air at constant flow rate, and (ii) degradation of hydrogen peroxide with catalase to release oxygen (section 2.3.2). In both experiments, the µ-Slide is placed on top of the microwave antenna on the stage of the microscope, and a region of the dried nanodiamond film was brought into the field of view through the antenna aperture with a 40× (0.6 NA) objective. A weight was placed on top of the slide (away from the excitation spot) for mechanical stability throughout measurements. Gas flow to the sample was provided via a silicone tube attached to one side of the slide while the other side was left open to atmosphere. **Figure 1A** shows a schematic of this experimental configuration. Flow was controlled by Brooks 5850E mass flow controllers (MFCs) (Brooks Instrument, Hatfield, PA, USA). The MFCs were ranged for 200 mL/min flow, with an accuracy of ± 1% of the full range (± 2 mL/min) and a repeatability of ± 0.25% for a given flow rate. For the gas mixture experiments, a static tube mixer (3067K19, McMaster-Carr, Elmhurst, IL, USA) was placed in the gas stream prior to the sample to promote uniform mixing of gases. For mixtures of pure oxygen and pure nitrogen at a total flow rate (Q) of 200 mL/min, the percentage of $O_2$ in the gas mixture was calculated with **Eqn. 1**:

$$\%\ O_2 = 100 \times \frac{q_{O_2}}{Q} \tag{1}$$

where the $q_{O_2}$ is the flow rate of $O_2$. To achieve more accurate concentrations of dilute (< 10%) $O_2$, pure $O_2$ was replaced with air (≈ 21% $O_2$) and mixed with nitrogen. The percentage of $O_2$ for a given flow rate of air ($q_{\text{air}}$) is given by **Eqn. 2**:

$$\%\ O_2 = 21 \times \frac{q_{air}}{Q} \tag{2}$$

As the pressure drop along the flow cell is estimated to be 0.5% of atmospheric pressure (see Supplemental Information), atmospheric pressure at the sample was assumed, and the partial pressure of $O_2$ for a given gas composition was calculated as 760 mmHg × the fraction $O_2$ in the gas mixture. Three ranges of % $O_2$ were investigated: (a) high range (0-100% $O_2$), (b) intermediate range (0-21% $O_2$), and (c) low range (0-5% $O_2$). The sample and gas lines were purged with pure nitrogen flow for at least 1 h prior to data acquisition.

### *2.3.2 Catalase – Hydrogen Peroxide Experiment*

A schematic for the catalase oxygen generation experiment is shown in **Figure 5A**. Catalase was prepared at a concentration of 10 mg/mL in deionized water. Then, 1 mL of this 10 mg/mL suspension was diluted to 15 mL with deionized water in a 50 mL conical centrifuge tube (0.67 mg/mL working catalase concentration). Separately, a 0.3 % (w/w), 88 mM, $H_2O_2$ solution was drawn into a 60 mL syringe affixed to a NE1000 syringe pump (New Era Pump Systems Inc., East

Farmingdale, NY, USA). The $H_2O_2$ syringe is connected to the lid of the centrifuge tube with silicone tubing and a Leur lock adapter. The catalase suspension was purged for 1 h with pure nitrogen flow at 4 mL/min while also being stirred with a magnetic stir bar, ensuring deoxygenation of the enzyme suspension and purging residual oxygen remaining in the gas lines. This nitrogen flush was maintained to drive $O_2$ evolved from the reaction to the nanodiamond sample and prevent backflow of gases into the lines. A slower flow rate of $N_2$ was selected for this experiment as compared to the all-gas experiments to avoid overdilation of the evolved $O_2$ which could degrade sensitivity. When data recording is started, the $H_2O_2$ solution is added with volumes of 2 mL (176 µmol), 4 mL (353 µmol), 6 mL (529 µmol), and 10 mL (882 µmol) at timepoints of 300 s, 1000 s, 1700 s, and 2400 s. Each addition step takes 60 s.

### 2.4 ODMR Measurements

#### 2.4.1 CW-ODMR Spectra and Contrast

CW-ODMR spectra were recorded at zero field and room temperature with ≈ 0.5 W incident microwave power as measured with a DK-3550 power meter (COMM-Connect A/S, Slangerup, Denmark). The recorded spectra were normalized by dividing each spectrum by the baseline amplitude, which was estimated by averaging the last 50 datapoints in each trace. ODMR contrast is defined as the $(PL_{baseline} - PL_{min}) / PL_{baseline} \times 100$, where $PL_{baseline}$ is the fluorescence baseline and $PL_{min}$ is the ODMR fluorescence dip minimum. Two-component Lorentzian fits of the $NV^-$ dips were used to derive the ODMR contrast directly.

#### 2.4.2 Modulated ODMR Contrast

A lock-in detection scheme with two modulated signals was used to record dynamic changes in the ODMR contrast. Experiments were run over one hour experiment intervals. In a typical measurement, the LED was modulated on and off at a frequency of 20 Hz and the microwave was simultaneously modulated on and off at a frequency of 120 Hz. Microwave were applied at 2.87 GHz, corresponding to the $NV^-$ zero field splitting resonance dip. A high microwave power of ≈ 0.5 W was used to maximize the ODMR contrast signal. The amplitudes of the modulated diamond fluorescence signals are extracted by demodulation performed entirely computationally. Data are collected every 4 s ($N = 900$ points for 1 h) by averaging of 8 oscilloscope traces, where each trace is 0.4 s long. The demodulated photoluminescence (PL) and microwave (MW) signals are ratioed (MW/PL) to extract the ODMR contrast, which is reported as a percent. A more detailed discussion on this method is given in the results section and further detailed technical description is provided in the Supplementary Information.

## 3. Results and Discussion

### 3.1 Gas Mixture Experiments

**Figure 1A** illustrates the experimental setup for gas mixture experiments. CW ODMR spectra are recorded sequentially as the percentage of $O_2$ in the gas mixture with $N_2$ is increased from 0% to 100% in 10% increments, corresponding to an $O_2$ partial pressure range from 0 to 760 mmHg. The major component of all the ODMR spectra (**Figure 1B**) is the $NV^-$ signal centered at 2.87

GHz. Another major dip corresponds to nearby P1 centers observed at 2.73 GHz with a symmetric lower intensity at just above 3 GHz. The asymmetry in the intensity of these two signals is a result of the microwave antenna being slightly detuned from 2.87 GHz. From the spectra, extracted CW ODMR contrast value are plotted as a function of oxygen partial pressure in **Figure 1C**. A raw linear coefficient of $(-7.02 \pm 0.33) \times 10^{-4}$ % $mmHg^{-1}$ $O_2$ is estimated from the linear fit. The CW ODMR contrast is an indication of the efficiency of $NV^-$ polarization; however, it also is significantly influenced by instrumental factors such as incident microwave and light power.[55] Because the absolute value of ODMR contrast could change depending on light delivery to the interrogated spot., the linear fit coefficient is normalized to the fitted intercept (4.53 ± 0.02 %, **Figure 1C**), providing a value that can be compared across different experimental runs. The fitted intercept is the estimated ODMR contrast at 0% $O_2$. This normalization results in a relative linear coefficient of $(-1.55 \pm 0.07) \times 10^{-4}$ % $mmHg^{-1}$. In this case, the slope represents the fractional change in ODMR contrast percent with $O_2$ partial pressure relative to a pure nitrogen calibration. The reduction in ODMR contrast is attributed to a diminished polarization of $NV^-$ centers due to paramagnetic oxygen collisions at the diamond surface that increase relaxation rates. In support of this conclusion, we observed that the NV $T_1$ relaxation rate increased with increasing $O_2$ concentration (**Figure S3**, Supplementary Information). The use of contrast as a means to measure paramagnetic analytes has been demonstrated,[25, 32] and is a means to correlate the relation of photoluminescence to relaxation rate.[56]

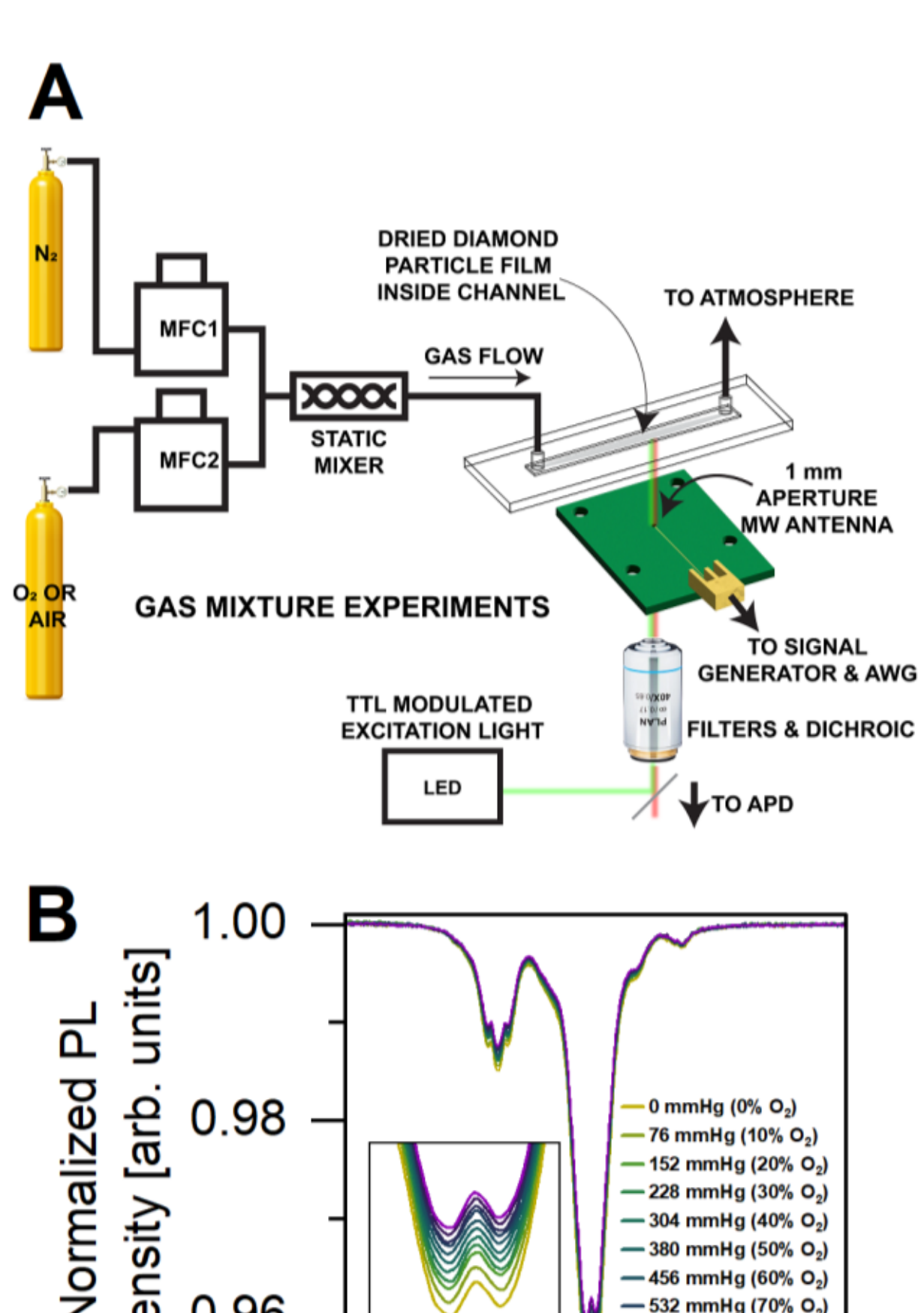
A
N₂
MFC1
MFC2
O₂ OR AIR
STATIC MIXER
GAS FLOW
DRIED DIAMOND PARTICLE FILM INSIDE CHANNEL
TO ATMOSPHERE
1 mm APERTURE MW ANTENNA
GAS MIXTURE EXPERIMENTS
TO SIGNAL GENERATOR & AWG
TTL MODULATED EXCITATION LIGHT
FILTERS & DICHROIC
LED
TO APD
B
Normalized PL Intensity [arb. units]
Microwave Frequency [GHz]
0 mmHg (0% O2)
76 mmHg (10% O2)
152 mmHg (20% O2)
228 mmHg (30% O2)
304 mmHg (40% O2)
380 mmHg (50% O2)
456 mmHg (60% O2)
532 mmHg (70% O2)
608 mmHg (80% O2)
684 mmHg (90% O2)
760 mmHg (100% O2)
υ [GHz]

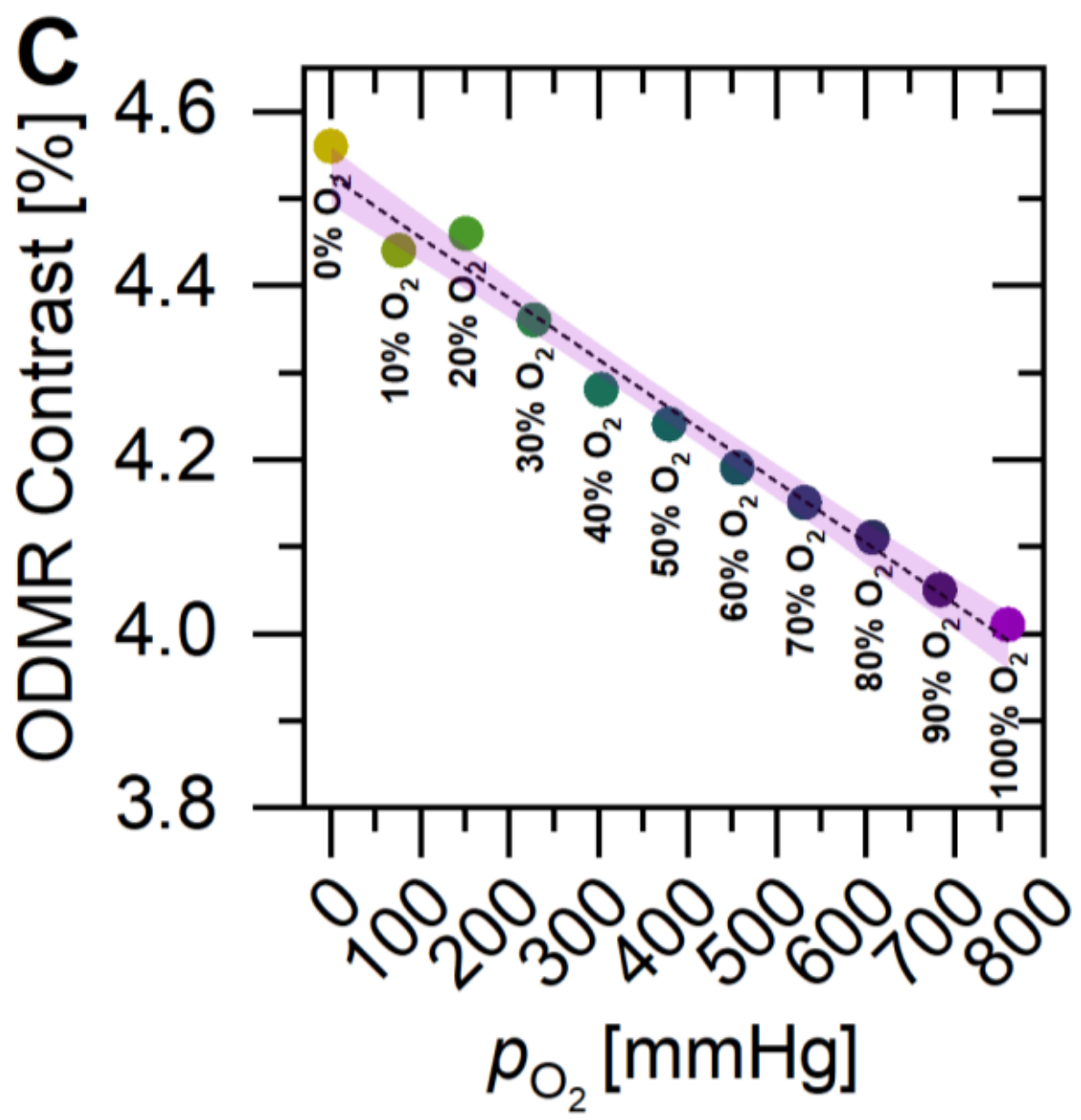
C
ODMR Contrast [%]
0% O2
10% O2
20% O2
30% O2
40% O2
50% O2
60% O2
70% O2
80% O2
90% O2
100% O2
pO2 [mmHg]

**Figure 1.** The influence of oxygen on CW ODMR spectra and contrast for the 70 nm diamond particles. **A**: Schematic of the gas mixture experiments. **B**: ODMR spectra collected at sequentially increasing levels of oxygen from 0% (pure $N_2$) to 100% (pure $O_2$). The inset shows a zoom-in of the 2.87 GHz $NV^-$ zero-field splitting dip. **C**: ODMR contrast as a function of oxygen partial pressure along with a linear fit to the data (dashed black line). The 95% confidence bands of the linear fit are indicated in magenta.

Motivated by these initial results with CW-ODMR, an alternative detection scheme was implemented using double modulation (double lock-in). Sarkar *et al.*[32] recently demonstrated the application of high frequency sinusoidal (harmonic) modulation of microwaves (~1 kHz) in combination with fluorescence modulation generated by the flow of microfluidic droplets containing 40 nm diamond particles at a droplet frequency ranging within *ca.* 30-350 Hz processed using harmonic Fourier analysis, including cross-frequency terms to isolate weak fluorescence signals. Our current approach utilizes square-wave modulation with a high duty-cycle (85%) fluorescence modulation generated by switching the LED excitation source on and off. An additional 120 Hz modulation of the microwave (50% duty cycle) is applied on top of the fluorescence modulation. An example of a single oscilloscope trace of this modulated signal is shown in **Figure S4** alongside a more detailed discussion of this method. These modulation conditions are relevant for slower relaxing, larger particles ($T_1 \approx 1$ ms for 70 particles in the dry state compared to 200-400 us in water). For our asymmetric square-wave scheme, harmonic decomposition becomes complex, producing both odd and even harmonics with cross-frequency interactions. Instead, the ODMR contrast is determined by a simple ratio of the microwave modulated fluorescence amplitude to the LED modulated fluorescence amplitude. Unlike the CW ODMR spectra, where quantities of interest (*e.g.*, the ODMR contrast) are derived from a two component Lorentzian fit of the full spectra, in the modulated ODMR experiments the ODMR contrast is generated in real-time from each component. As a result, the square-wave modulation scheme offers several advantages: (1) high precision measurements are possible due to noise rejection resulting from the lock-in methodology; (2) high duty cycle of light illumination (85%) increases the effective NV optical excitation beyond typical light chopping schemes (~50%) to increase signal; and (3) the ability to monitor dynamic analyte changes in real time.

**Figure 2** shows the results of the double modulation measurements for the different ranges of oxygen concentrations. **Table S1** in the Supplemental Information lists the experimental details of the measurements performed at each oxygen sweep range. In **Figure 2A**, $O_2$ partial pressure was changed from 0-100 %, with clearly visible steps resulting from a decrease in ODMR contrast as the $O_2$ content is incrementally increased. Averaged ODMR contrast values across the individual time windows of the steps are plotted as a function of $O_2$ partial pressure in **Figure 2B**, providing an intercept normalized slope of $(-1.73 \pm 0.05) \times 10^{-4}$ mmHg$^{-1}$, similar to the normalized relative slope determined from the CW-ODMR spectra (**Figure 1C**, $(-1.55 \pm 0.07) \times 10^{-4}$ mmHg$^{-1}$). From **Figure 2A**, the smallest detectable change in ODMR contrast is estimated to be $9 \times 10^{-3}$ %, calculated from 3σ of the given window average ($\sigma = 3\times10^{-3}$ %). This change in ODMR contrast corresponds to a projected minimum detectable differential change in $O_2$ partial pressure of ≈ 9 mmHg as estimated from the raw slope of $(-10.1 \pm 0.3) \times 10^{-4}$ % $mmHg^{-1}$ (**Table 1**).

To validate this projected limit of detection, two lower oxygen concentration ranges were investigated, as shown in **Figure 2**. In these experiments, after sequential addition of $O_2$, to the highest amount, either 21 % or 5 %, the $O_2$ content was then sequentially decreased to assess measurement repeatability and hysteresis, indicated by the red portions and datapoints in **Figure 2**. The raw and normalized slopes average derived ODMR contrast values from **Figure 2D** and **Figure 2F** are summarized in **Table 1**. Notably, in the lowest range (0-5% $O_2$), returning to lower $O_2$ conditions after exposure to higher $O_2$ conditions leads to slightly reduced ODMR contrast values, leading to differing slopes (**Figure 2D** and **Figure 2F**). This hysteresis may be a result of insufficient removal of physiosorbed $O_2$. In these lower concentration ranges, the 3σ value remains the same at $9\times10^{-3}$ % ODMR contrast. The resulting projected minimum detectable differential change in $O_2$ partial pressure are summarized in **Table 1**. These estimated detection limits are lower than the 9 mmHg estimated from the high $O_2$ range due to the different slopes, but are still in reasonable agreement. Thus, the projected limit of detection estimated from the full $O_2$ sweep range (**Figure 2A** and **2B**) is experimentally confirmed at low $O_2$ sweep ranges to the order of about $\Delta pO_2 \approx 6.5$ mmHg (taking the average of all of the estimated limits of detection). A possible source for these different slopes is an additional mechanistic process happening when the nitrogen saturated surfaces are disrupted with oxygen. When switching from pure $N_2$ to $N_2/O_2$ (or $N_2$/Air) mixtures, there is an initially larger and slower drop in the ODMR contrast (**Figure 2A**, **2C**, **2E**). The ODMR contrast can then take minutes to stabilize. For example, when the gas was switched from pure $N_2$ to 10% of $O_2$, (**Figure 2A**), the ODMR contrast requires *ca*. 2 min to stabilize, whereas for lower $O_2$ additions (*e.g.*, 2.1% **Figure 2C**, 1.1% **Figure 2E**), the ODMR contrast stabilizes over *ca*.10 min. While the higher percentage of $O_2$ in the pure $O_2/N_2$ mixtures as compared to Air/$N_2$ should saturate the diamond particle surfaces faster, a slow re-equilibration process is still observed after the initial addition of $O_2$. Even in the case of the ODMR contrasts derived from the ODMR spectra (**Figure 1C**), there is apparently a less linear regime in the lower $O_2$ range (<20%). Overall, measuring at lower $O_2$ sweep ranges effectively zooms in on this secondary regime, leading to different slopes even after normalization. More precise control of the $O_2$ content is needed to explore this transition between pure $N_2$ and $O_2$ in detail. The repeatability of the measurements can be estimated by calculating the standard deviation of ODMR contrasts values upon repeated measurements at the same concentrations. We calculate a repeatability of approximately 0.006 in percent ODMR contrast.

**Table 1.** Summarized Raw and Normalized ODMR Contrast % Dependencies to $O_2$

| **ODMR** | **$O_2$ Range [%]** | **Raw Slope [% mmHg$^{-1}$]** | **Intercept Normalized Slope [ mmHg$^{-1}$]** | **Minimum Detectable $\Delta pO_2$ [mmHg]** |
|---|---|---|---|---|
| CW | 0→100 | $(-7.02 \pm 0.33) \times 10^{-4}$ | $(-1.55 \pm 0.07) \times 10^{-4}$ | N/A |
| Modulated | 0→100 | $(-10.1 \pm 0.3) \times 10^{-4}$ | $(-1.73 \pm 0.05) \times 10^{-4}$ | 9 |
| Modulated | 0→21 | $(-14.9 \pm 1.0) \times 10^{-4}$ | $(-2.53 \pm 0.17) \times 10^{-4}$ | 6 |
| Modulated | 0→5 | $(-22.4 \pm 2.6) \times 10^{-4}$ | $(-3.81 \pm 0.44) \times 10^{-4}$ | 4 |
| Modulated | 5→1.1 | $(-12.4 \pm 0.3) \times 10^{-4}$ | $(-2.12 \pm 0.06) \times 10^{-4}$ | 7 |

In addition to altering ODMR contrast, molecular $O_2$ also quenches the diamond fluorescence signal (**Figure S6**). Oxygen induced fluorescence quenching is a well-known phenomenon[57] which is also exploited for oximetry in fluorescence based oxygen sensors.[58, 59] For simple collisional quenching, the relationship between the fluorescence and concentration follows the *Stern-Volmer* equation[60] (**Eqn. 3**):

$$\frac{F_0}{F} = 1 + K_{SV} p_{O_2} \quad (3)$$

Where $F_0$ and $F$ are the fluorescence in the absence and presence of a quencher, $K_{SV}$ is the quenching constant, and $p_{O_2}$ is the oxygen partial pressure. We consider only the high $O_2$ sweep range (**Figure 2A**) for this analysis. The values of $F$ are estimated by averaging the PL intensities within each $O_2$ concentration region of interest (ROI), with $F_0$ being taken from the average fluorescence under the pure $N_2$ flow. The fluorescence signals used for this analysis are shown in **Figure S5A** in the Supplemental Information. **Figure 3** shows the *Stern-Volmer* plot of the high $O_2$ sweep range. A non-linear trend is observed, which is typical when there are fractions of accessible and non-accessible fluorophores[60]; however, as saturation is not reached even at 1 atm (760 mmHg), experiments at elevated $O_2$ pressures would be desirable in the future. Typically, $O_2$ quenches fluorescence by interaction with the excited state of a fluorophore, requiring essentially direct contact between the fluorophore and $O_2$ molecules. An alternative mechanistic consideration of the observed behavior could be a charge shift in the $NV^-/NV^0$ ratio caused by physiosorbed $O_2$. However, the NV PL spectra measured under either pure $O_2$ or $N_2$ flow does not indicate any increase or decrease in the $NV^0$ zero phonon line (ZPL) at 575 nm with respect to the $NV^-$ ZPL at 637 nm, and only a slight drop in the $NV^-$ luminescence is seen (**Figure S6**, Supplementary Information). Kumar *et al.*[49] suggested that $O_2$ can lead to an increase in $NV^-$ fraction, as compared experiments in an oxygen atmosphere with those in vacuum. When investigating the role nitrogen terminated groups on the surface, Kawai *et al.*[61] concluded that the fully nitrogen terminated diamond stabilizes the near surface $NV^-$ centers due to a high positive electron affinity. Our results suggest that physiosorbed nitrogen (not covalently bound) promotes higher ODMR contrast as compared to $O_2$ containing environments down to at least *ca.* 1% $O_2$. Given the recent works assessing the changes of energy band structures in response to surface functionalization[61-63] or in vacuum versus air[49], our results indicate that further assessment on the role of specific gas environments on $NV^-$ fluorescence and charge state is needed.

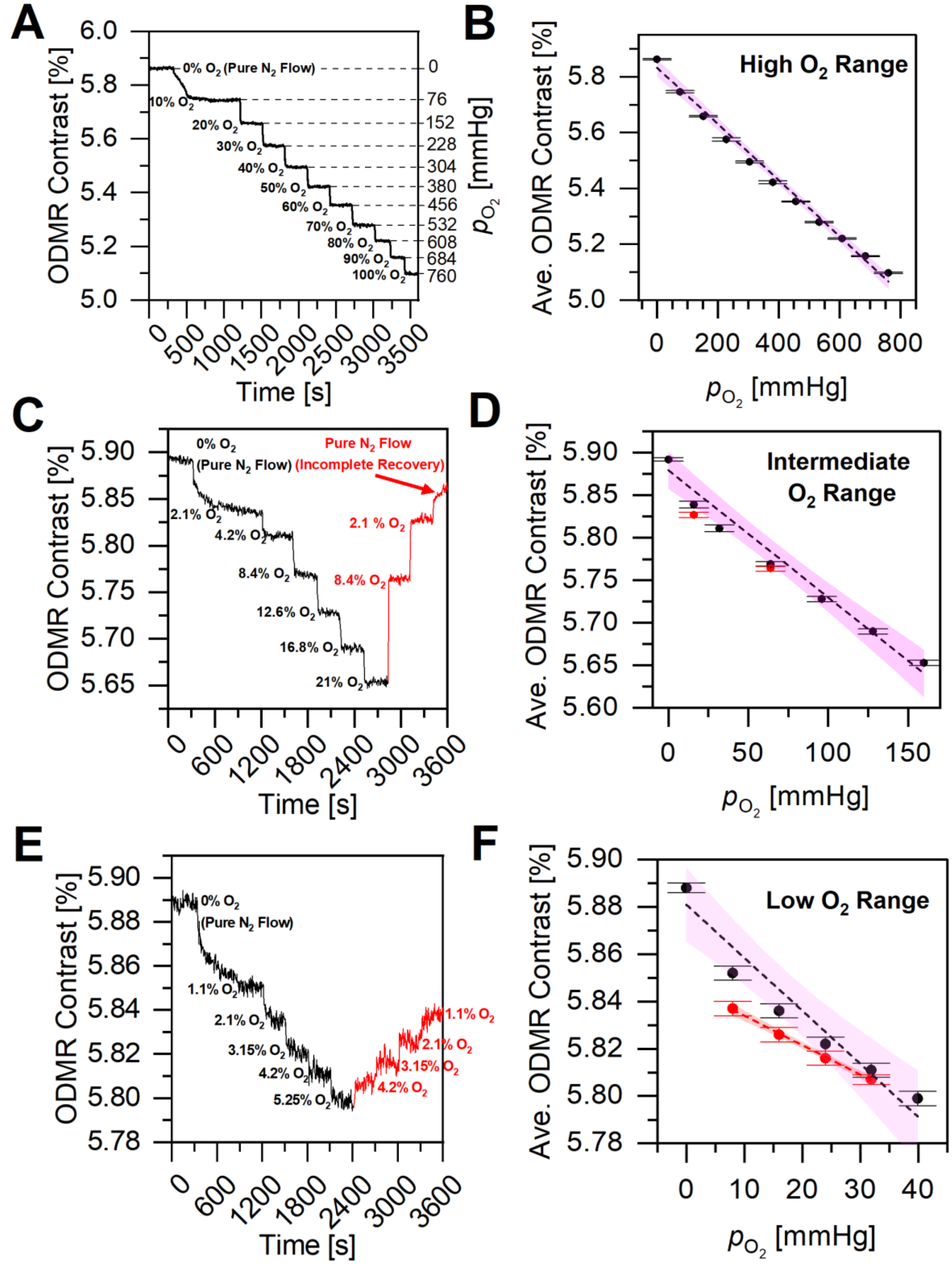

**Figure 2.** ODMR contrast measured from modulated ODMR data and their corresponding ratio for a high $O_2$ sweep range (0-100%) **A** and **B**, intermediate $O_2$ sweep range (0-21%), **C** and **D**, and low $O_2$ sweep range (0-5%), **E** and **F**. For the intermediate and low sweep ranges, the $O_2$ content was increased to a maximum (black traces and circles), and then reduced back to lower amounts (red traces and circles) to monitor recovery of the signals as $O_2$ was removed.

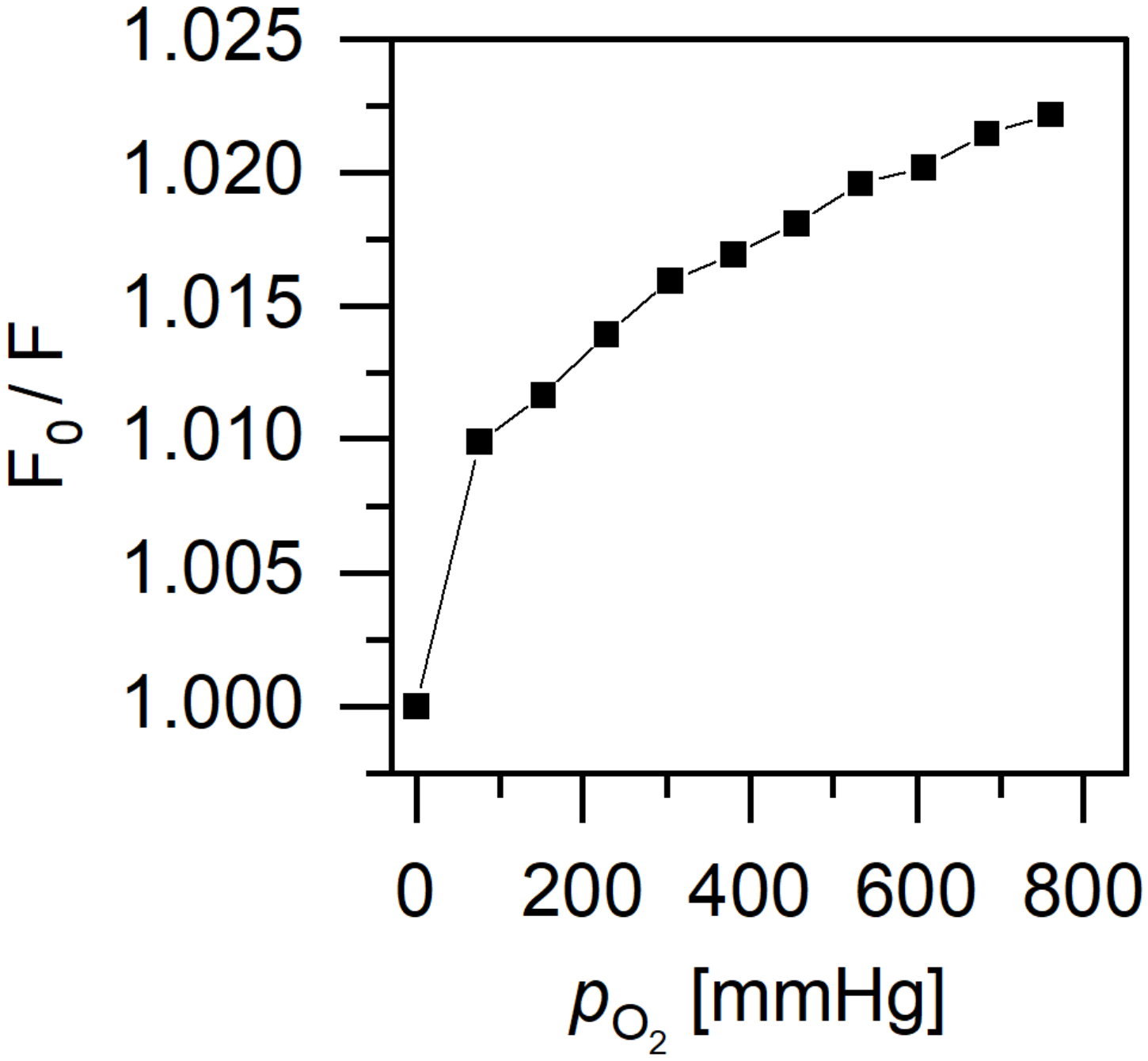

**Figure 3.** Stern-Volmer plot for 70 nm fluorescent diamond particles when oxygen partial pressure was increased from 0 mmHg (pure nitrogen) to 760 mmHg (pure oxygen) at ambient pressure.

### 3.2 Catalase Experiments

The potential for detection of $O_2$ by NV quantum sensing was explored by detecting $O_2$ generated from a model reaction – a catalytic degradation of hydrogen peroxide in the presence of the catalase enzyme:

$$2\ H_2O_2 \rightarrow 2\ H_2O + O_2$$

Catalase (E.C. 1.11.1.6) proteins are a widely studied enzyme family that protect cells against oxidative stress by degrading $H_2O_2$ before it can form potentially harmful reactive oxygen species (ROS).[64, 65] The catalase employed in this study was derived from *Aspergillus niger* (*A. niger*) and contains four heme subunits.[66] The experimental setup is illustrated in **Figure 4A**. Specified amounts of $H_2O_2$ are added at times indicated by the red dashed lines on **Figure 4B**. A corresponding drop in the ODMR contrast follows as the $O_2$ is evolved. To assess the magnitude of this response from **Figure 4B**, the mean baseline ($\mu_{baseline}$) is determined from the mean value of the first 75 and last 51 datapoints of the trace, while mean signal amplitude ($\mu_{signal}$) for each addition of $H_2O_2$ is estimated by taking the mean of 33 datapoints near the bottom of each of the dips that are observed after addition of $H_2O_2$. Because generated $pO_2$ is not known *a priori*, the response magnitude ($\Delta\mu = |\mu_{baseline}-\mu_{signal}|$) is plotted against moles of $H_2O_2$ added, indicating a linear relationship with a slope of $(1.51 \pm 0.13) \times 10^{-4}$ % $\mu mol^{-1}$ $H_2O_2$ added (**Figure 4C**). The noise level of the baseline ($\sigma_{baseline}$) is estimated to be 0.0054 %, implying the 3σ detection limit is the amount of evolved oxygen from 107 μmol of $H_2O_2$. Practically, this means in the current experimental configuration, the burst of $O_2$ produced from our lowest addition (176 μmol $H_2O_2$ added) is near but above our detection limit.

From **Figure 4C**, the observed reduction of ODMR contrast of 0.015 % from the addition of 176 μmol $H_2O$ correlates to a measured $\Delta pO_2$ of ≈ 15 mmHg using the slope of $(-10.1 \pm 0.3) \times 10^{-4}$ % $mmHg^{-1}$ (**Figure 2B**). To relate this to the generated oxygen, the total activity of catalase in the reaction vessel was 2,000 μmol $min^{-1}$, thus 176 μmol of $H_2O_2$ would be consumed in ≈ 5.3 s producing 88 μmol $O_2$. This gives an approximate $O_2$ formation rate of 996 μmol $O_2$ $min^{-1}$. For a fixed volume of 0.033 L, the change in $O_2$ partial pressure is given by **Eqn. 4**,

$$\frac{dpO_2}{dt} = \frac{RT}{V}\frac{dnO_2}{dt} \quad (4)$$

resulting in a formation rate of $O_2$ of ≈ 565 mmHg $min^{-1}$, or ≈50 mmHg for the given 5.3 s burst. Given that nitrogen flow dilutes the amount of oxygen somewhat, a measured $\Delta pO_2$ of ≈ 15 mmHg is reasonable.

For sampling, there is a tradeoff between the achievable sensitivity, which would be limited by the dilution factor of the formed $O_2$ in the $N_2$ stream, and the time resolution (e.g., if the goal was to monitor reaction kinetics). A higher $N_2$ flow rate would quickly sweep the evolved $O_2$ to the sample, facilitating faster detection; however, the sensitivity could be reduced to the higher dilution of $O_2$. For an assumed 88 μmol of produced $O_2$, this corresponds to 2.2 mL $O_2$ over the ~5.3 s, giving a volumetric $O_2$ production rate of 24.9 mL $min^{-1}$. Since the $N_2$ flow rate (4 mL $min^{-1}$) is much less than this production rate, this should put us in a higher sensitivity regime, since flow rates higher than the production rate would dilute the evolved $O_2$ faster. Our detection timescale is much slower than the actual reaction timescale, with a minimum in ODMR contrast occurring about 300 s after $H_2O_2$ addition (**Figure 4B**).

These results indicate a linear response of NV quantum sensor to amount of $H_2O_2$ added, and, correspondingly, the amount of $O_2$ generated. A better understanding of the adsorption/desorption processes at the diamond surfaces could assist in further result interpretation. A next generation diamond sensor design could, for example, confine all particles in a narrower constriction, such that all evolved $O_2$ could come into contact with a larger fraction of the particles under measurement. A more uniform distribution of particles would also benefit a sensor platform so that particles are uniformly distributed across the sampling area.[67] The use of smaller particles such as 40 nm may further improve sensitivity.

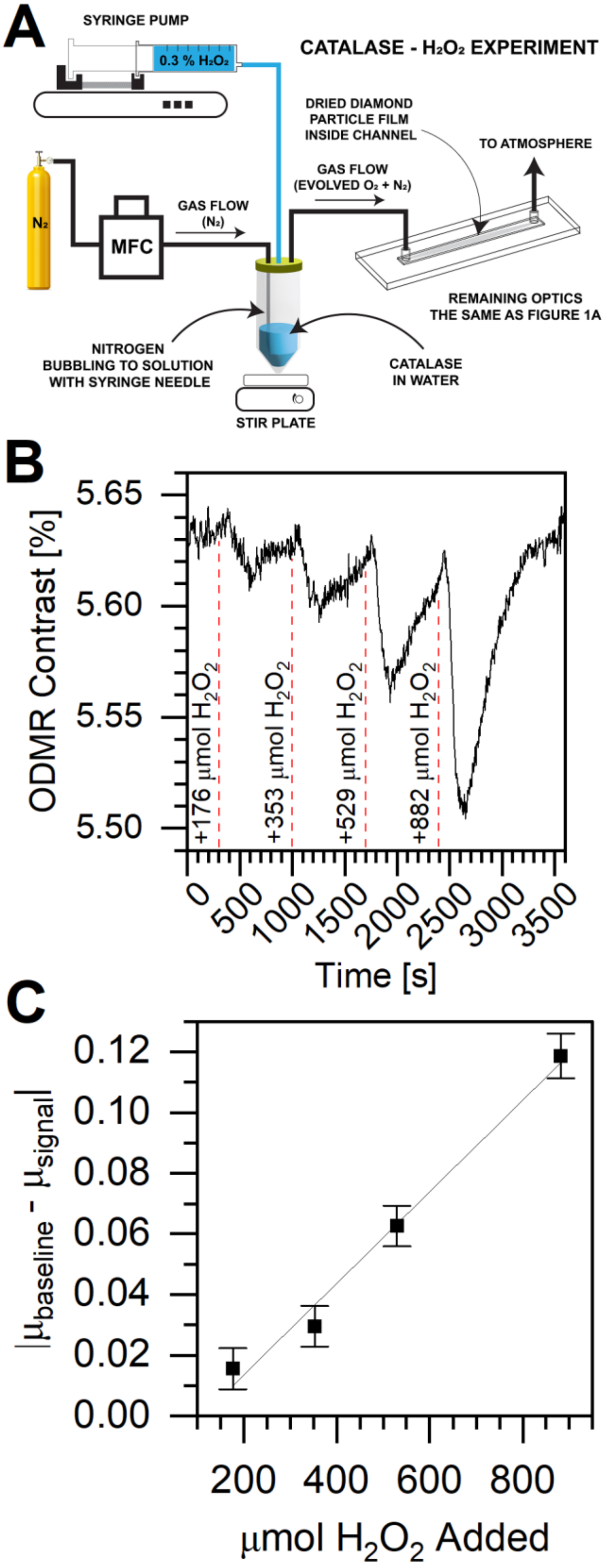

A
SYRINGE PUMP
0.3 % H₂O₂
CATALASE - H₂O₂ EXPERIMENT
DRIED DIAMOND PARTICLE FILM INSIDE CHANNEL
TO ATMOSPHERE
GAS FLOW (EVOLVED O₂ + N₂)
N₂
MFC
GAS FLOW (N₂)
REMAINING OPTICS THE SAME AS FIGURE 1A
NITROGEN BUBBLING TO SOLUTION WITH SYRINGE NEEDLE
CATALASE IN WATER
STIR PLATE
B
ODMR Contrast [%]
5.65
5.60
5.55
5.50
+176 µmol H2O2
+353 µmol H2O2
+529 µmol H2O2
+882 µmol H2O2
0
500
1000
1500
2000
2500
3000
3500
Time [s]
C
|µbaseline - µsignal|
0.12
0.10
0.08
0.06
0.04
0.02
0.00
200
400
600
800
µmol H2O2 Added

**Figure 4.** Detection of $O_2$ evolved from the catalytic degradation of $H_2O_2$. **A**: Schematic of the experimental setup. **B**: Modulated ODMR signal collected over 1 h with additions of known quantities of $H_2O_2$ indicated by the red dashed lines. **C**: Magnitudes of the change in ODMR contrast determined from **B** as a function of added amounts of $H_2O_2$ with a linear fit (black line).

## 4. Conclusion and Outlook

The current study reports on the use of fluorescent nanodiamond particles for the detection of paramagnetic molecular $O_2$ in the gas phase — to the best of our knowledge, a first publication on the topic. There is a significant body of literature that reports on the use of NV centers in diamond for the detection of paramagnetic species such as reactive oxygen species and metal ions such as $Gd^{3+}$. However, the use of NV centers to detect paramagnetic $O_2$ is surprisingly scarce despite that $O_2$ is one of the most well-studied and ubiquitous paramagnetic relaxers and a strong quencher of fluorescence. The effect of dissolved $O_2$ on the relaxation of NV centers was studied in aqueous solutions of NV-containing diamond nanoparticles[41, 53] and in a microfluidic setup using a bulk diamond plate with shallow implanted NV centers.[50] There is controversy on the ability to resolve dissolved $O_2$ using NV relaxometry (positive for a diamond plate with shallows implanted NVs and negative for nanodiamond particles). In the gas phase, Kumar, R. et. al. observed prolonged $T_1$ of NV centers in air in comparison to a vacuum environment.[49] In the current work, we report a quantitative detection of gaseous molecular $O_2$ that is flowed over a film of dried diamond particles. We observed linear changes of ODMR contrast and $T_1$ relaxation rate with changes in $O_2$ partial pressure with capability to resolve 1% of $O_2$ in a mixture with $N_2$ using modulated ODMR contrast measurements. Overall, these novel results emphasize importance of effects of physisorbed $O_2$ on the near surface NV centers, guiding future fundamental studies related to electronic structures of near surface NV centers.[61, 63] In a practical sense, surface contamination from molecular $O_2$ could be a source of a degraded performance of single NV diamond scanning probes. Moreover, the reported results on the gas phase $O_2$ detection open new directions for NV based sensing platforms, owing to emerging NV sensing technologies providing (sub)micron spatial resolution on distribution of physical parameters (magnetic fields, temperature, ROS) within a sample. Similarly, using diamond nanosensors, spatial imaging of $O_2$ at a (sub)micron scale can address unanswered questions related to the heterogeneity of local oxygen, indistinguishable via bulk sensing.[68] Applications areas are wide ranging. There are needs in catalysis where local gradients can affect performance yet are challenging to assess.[69, 70] In micro-electro-mechanical systems (MEMS), oxygen can locally contribute to device failure.[71-73] In biology, nitrogen fixation is essential for life, yet the ability of cyanobacteria to coordinate nitrogen fixation with oxygen photosynthesis is still not well understood.[74] Resolving such photosynthetic spatiotemporal mechanisms could enable engineering of cyanobacteria to improve fixation efficiency.

## Acknowledgements

Research reported in this publication was supported by the National Center for Advancing Translational Science of the National Institutes of Health under award number R44TR005336. The content is solely the responsibility of the authors and does not necessarily represent the official views of the National Institutes of Health.

## Conflict of Interest

N.N., A.M., O.S., and M.T. are employees of Adámas Nanotechnologies, Inc. A for-profit commercial entity engaged in the sale of diamond materials and quantum instrumentation.

**Supplementary Information**

# Quantitative Detection of Molecular Oxygen in the Gas Phase with Fluorescent Nanodiamonds

Nicholas A. Nunn*[1], Antonin Marek[1], Alex I. Smirnov[2], Olga A. Shenderova[1], Marco D. Torelli*[1]

[1]Adámas Nanotechnologies, Inc., Raleigh, NC, USA

[2]Department of Chemistry, North Carolina State University, Raleigh, NC, USA

*Corresponding Authors:

nnunn@adamasnano.com

mtorelli@adamasnano.com

## S1. Calculation of Pressure Drop Across Channel of μSlide

If atmospheric pressure is assumed (760 mmHg), then the $O_2$ partial pressure can be estimated as 760 × fraction of $O_2$ in gas mixture. The pressure drop across the slide channel is estimated to be negligible according to the Hagen-Poiseuille relationship for a rectangular channel with $h << w$ (**Eqn. S1, S2**)[1]:

$$\Delta P \approx \mathrm{Q} \cdot \mathrm{R_h} \qquad \textbf{(S1)}$$

$$\mathrm{R_h} = \frac{12\mu \cdot \mathrm{L}}{\mathrm{w} \cdot \mathrm{h}^3} \qquad \textbf{(S2)}$$

Where $R_h$ is the hydraulic pressure and Q is the flow rate. $\mu$ is the dynamic viscosity at 25 °C – approximately $1.85 \times 10^{-5}$ Pa-s for air. $L$, $w$, and $h$ are the channel length ($5\times10^{-2}$ m), channel width ($5\times10^{-3}$ m), and channel height ($2.5\times10^{-4}$ m), correspondingly. Q, the flow rate, is 200 mL/min ($3.33 \times 10^{-6}$ m/s). Thus, a pressure drop of about 474 Pa is estimated, which is only about 0.5% of atmospheric pressure (101,325 Pa); hence, we assume that the gas is near atmospheric pressure for oxygen partial pressure estimations.

## S2. Example of a CW ODMR Spectrum with a Least-squares Fit

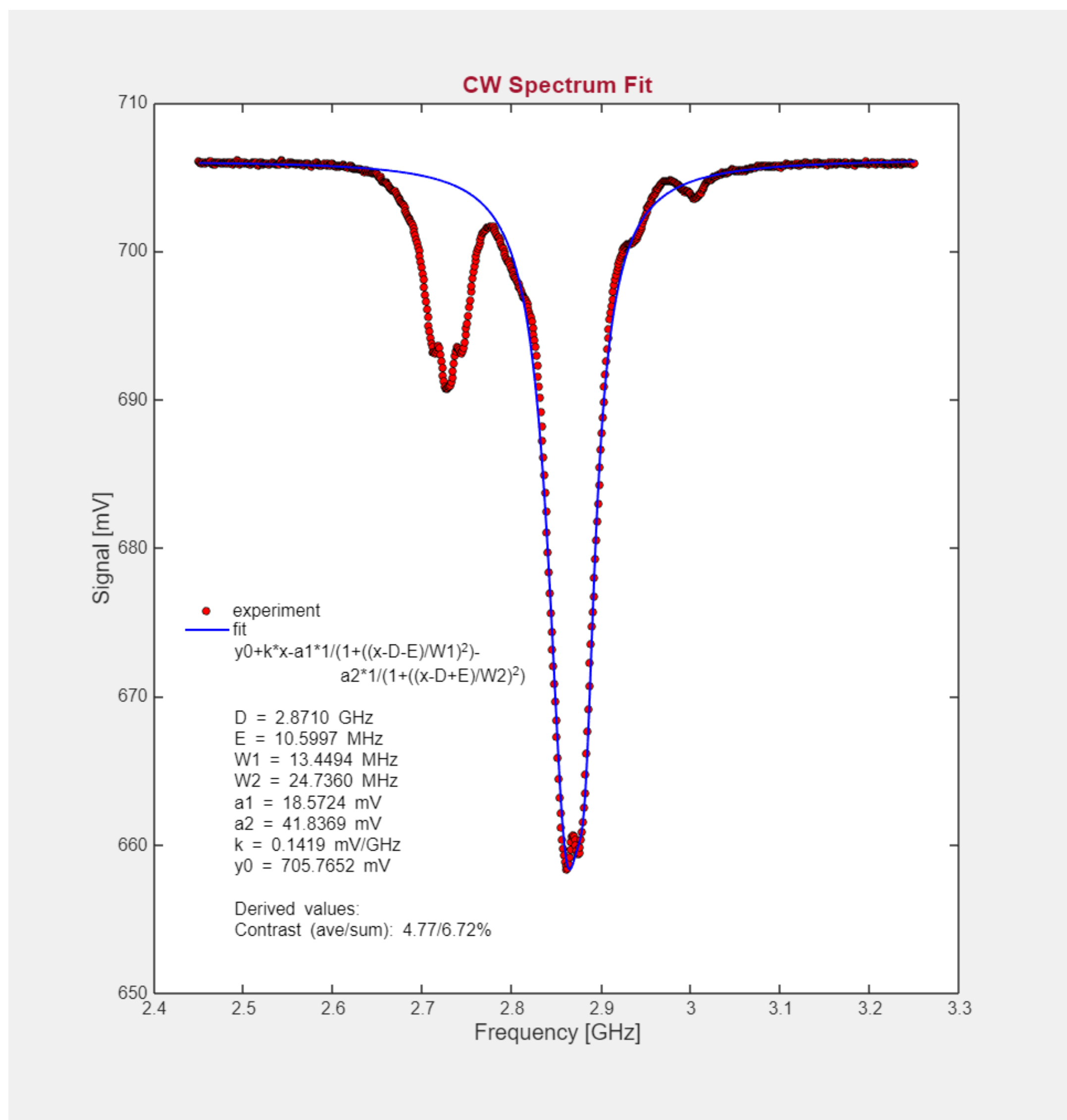

**Figure S2.** Example of a CW ODMR spectrum collected for the 70 nm particles showing fit parameters. For the ODMR contrast reported in the main text, we refer to the sum value (6.72% in this case), which is the contrast for the sum of the two Lorentzian functions.

## S3. NV $T_1$ Relaxation Rate Versus Oxygen

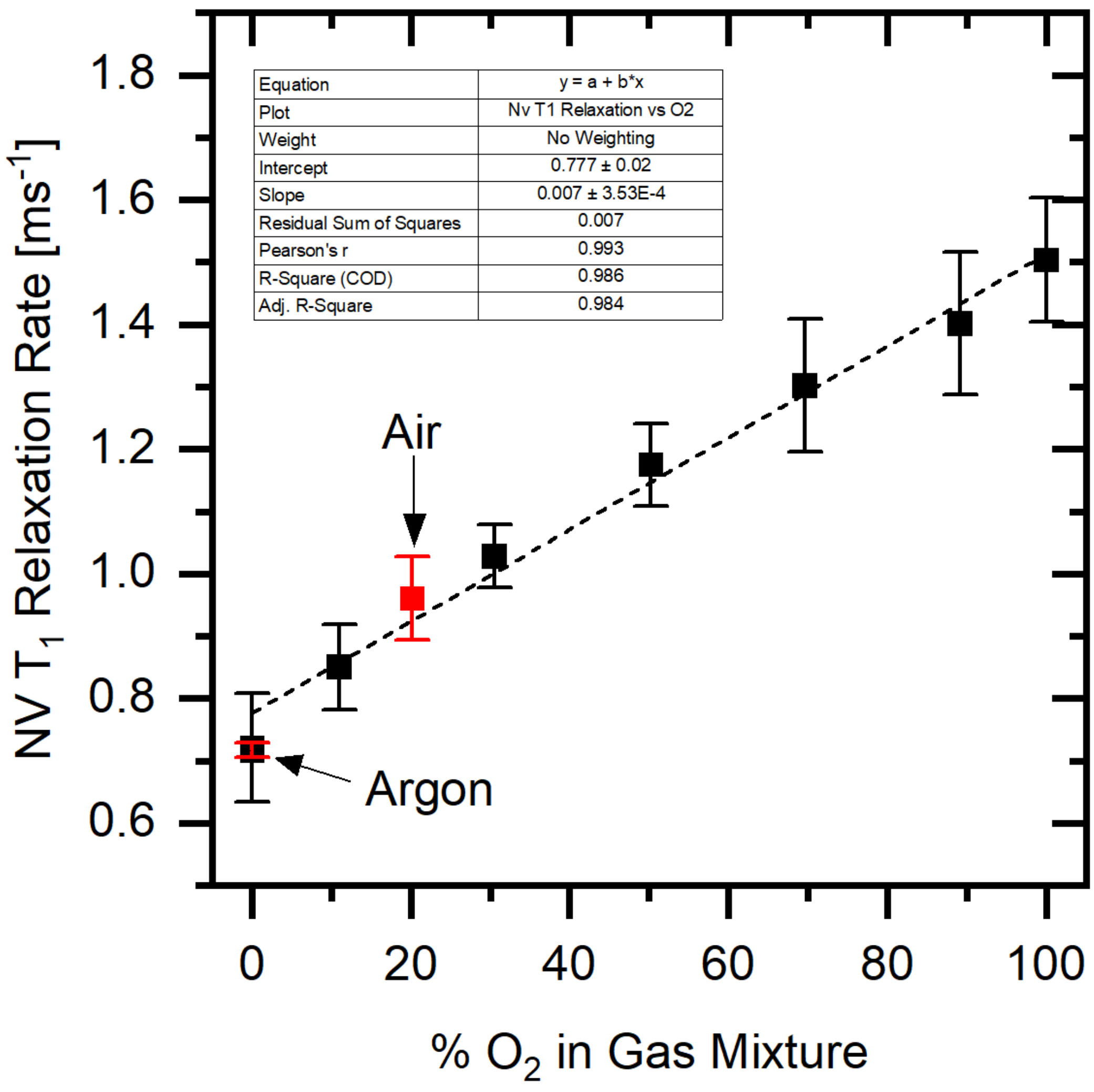

**Figure S3.** NV $T_1$ relaxation rates of dry 70 nm particles measured as a function oxygen content in a mixture with nitrogen gas. Repeated measurements of $T_1$ relaxation traces are globally fit with a stretched exponential function with shared stretching parameter. The relaxation rate is calculated as $1/T_1$. A linear fit to the extracted relaxation rates is shown as the dashed line. Measurements performed in air and argon are indicated in red.

## S4. Double Modulation Methodology

### ODMR setup

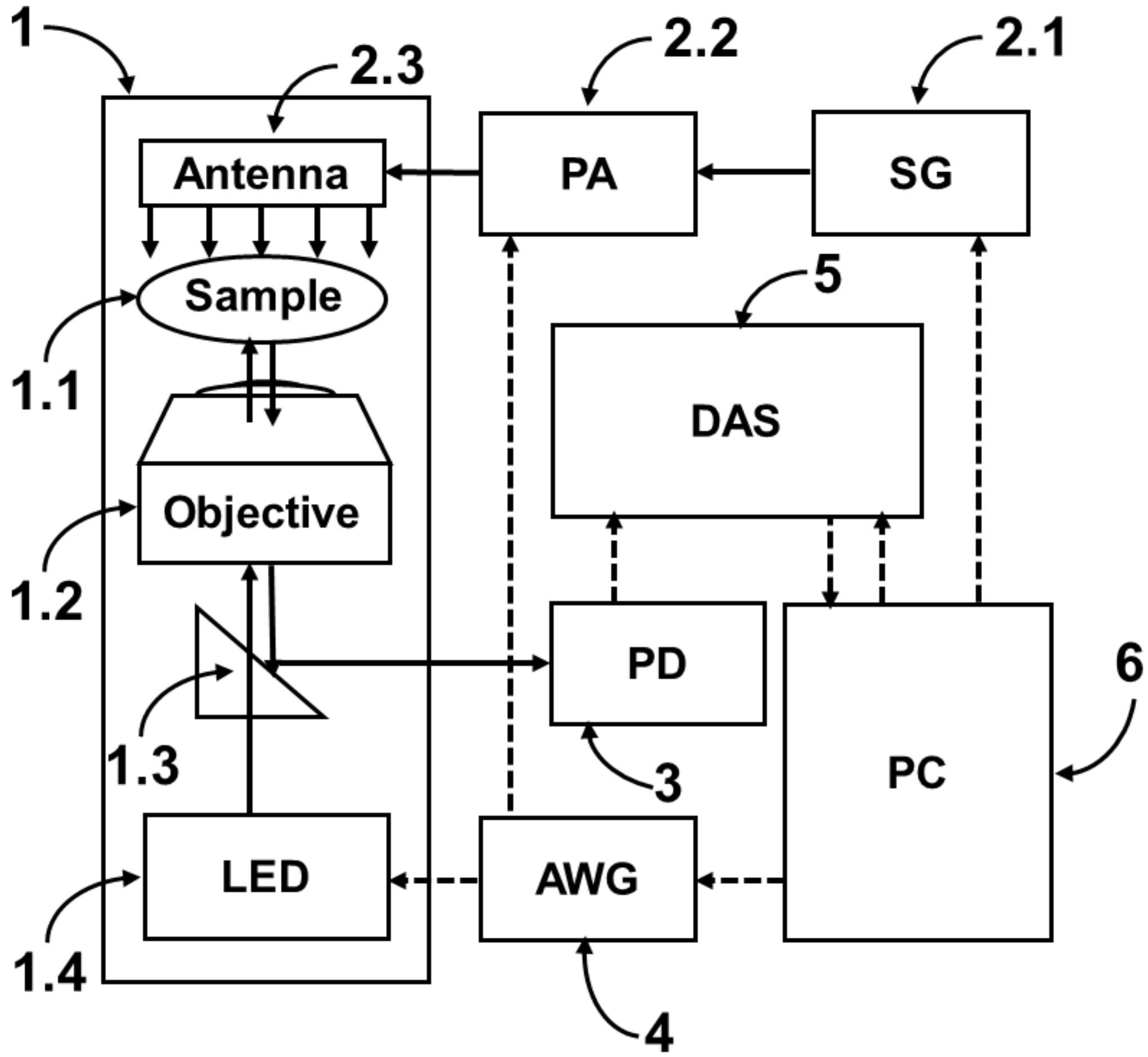

**Scheme 1.** A block diagram of the ODMR set-up.

The ODMR-Pulse.v1 system is composed of several key instruments, see **Scheme 1**. Major optical components are centered around a **fluorescence microscope** (1). The **sample** (1.1) is positioned on a mechanical stage capable of movement in both the XY (horizontal) and Z (vertical) directions. Through a selection of **objectives** (1.2), typically 10× and 40×, the excitation light from an **LED source** (1.4) is focused onto the sample. After the excitation, emitted fluorescence light is collected through a set of **filters and mirrors** (1.3) and directed towards the **photodetector** (PD, 3). Microwave generation starts with a **signal generator** (SG, 2.1). Microwaves are then transmitted through various components such as **power amplifier** (PA), a switch, a mixer, and a filter (2.2). Ultimately, the microwaves are delivered onto the sample by an **antenna** (2.3). The propagation of light and microwaves is represented by solid lines in the figure (—). The experiment is interfaced and controlled by a **PC** running a custom **ADAMAS ODMR** software (6). The program communicates with LED and SG via an **arbitrary waveform generator** (AWG, 4). The AWG is used as a programmable modulation source and TTL logic device capable of synchronously manipulating light and microwaves, hence precisely controlling timing and duration of modulation sweeps and pulses. The analog signal from a photodetector (PD) is collected by a data acquisition system (DAS, 5), an **oscilloscope** in our case. Subsequently, the processed signal is transferred back to a PC/AODMR for evaluation. The signal pathways (command, logic, and data) are represented by dashed lines (---).

### 1) Modulation ODMR principles

Modulation ODMR experiment is designed to provide pulse-modulated input signals while simultaneously demodulating the resulting output signal amplitudes. It can handle up to two independent modulation sources at the same time. One source always modulates the excitation light source, which can be further combined with either microwave source modulation at the resonance frequency, as in this case, or other external sources, such as a magnetic field. The experiment allows for real-time tracking of the demodulated amplitudes from one or both modulation sources, providing live updates at a fast rate (typically up to 4 samples per second). The two modulation inputs are generated using the two available channels of the AWG, ensuring stable and synchronized timing over extended periods. The demodulation is performed entirely in software, without the use of dedicated hardware lock-in detection. The averaged signal from the oscilloscope is combined with the known modulation input and processed to extract the output signal amplitude corresponding to each modulation source. This purely computational approach minimizes expensive hardware dependencies while maintaining high accuracy and reliability.

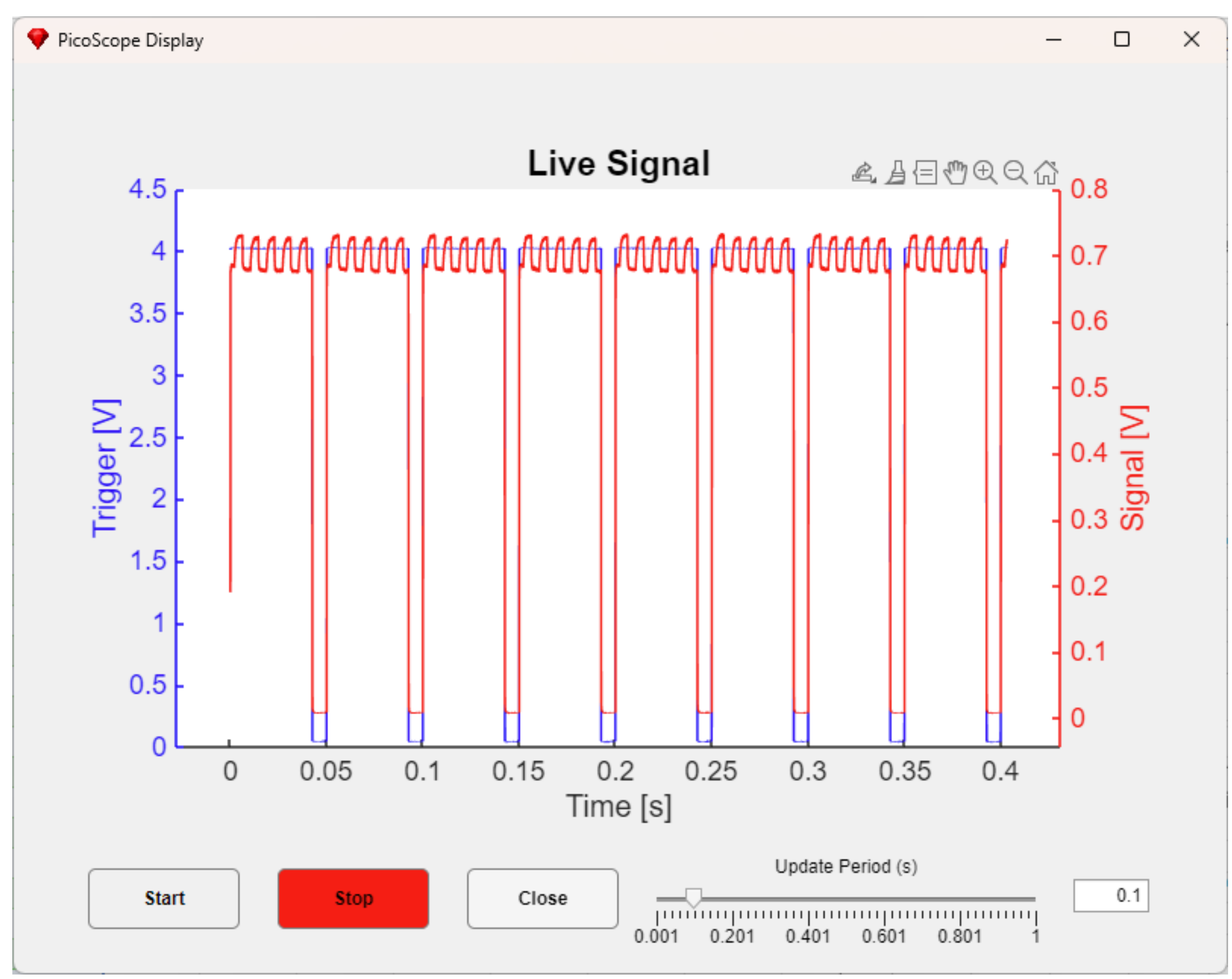

**Figure S4:** A representative time trace of the doubly modulated ODMR sample signal (red line). The optical and microwave excitations are modulated at distinct, well-defined frequencies, and the detected sample response follows this combined modulation pattern. The lower optical modulation frequency (20 Hz) is apparent from the periodic intervals during which the sample signal rises and falls to zero. Superimposed on the light-on intervals, the higher microwave modulation frequency (120 Hz) manifests as a weaker periodic variation. The optical excitation is operated at a high duty cycle of 85%, whereas the microwave excitation employs a symmetric 50% duty cycle. As a rule, the lower-frequency excitation modulation—here the optical modulation—also serves as the oscilloscope trigger signal (blue line). This approach ensures consistent phase alignment during signal averaging and directly illustrates the asymmetric duty cycle in phase with the sample response.

In principle, the amplitude of a modulated sample signal can be efficiently extracted using a known reference modulation waveform via digital lock-in detection. The method relies on the orthogonality of harmonic functions. By multiplying the detected sample signal with the reference signal and averaging over a sufficiently long time interval (i.e., over multiple modulation periods), one obtains a value proportional to the signal amplitude at the reference frequency. For sinusoidal modulation, the averaged value equals one-half of the signal amplitude, provided that the reference and signal are in phase. In oscilloscope-based acquisition, in-phase detection can be readily achieved through deterministic triggering. For doubly modulated signals, the trigger must be derived from the lower modulation frequency to preserve a stable phase reference and ensure consistent averaging.

These considerations extend naturally to doubly modulated ODMR with square-wave excitation. While phase coherence is maintained via deterministic triggering, symmetric optical modulation (50% duty cycle) is inefficient: half of the acquisition time corresponds to zero signal, limiting sensitivity to the weaker microwave-induced modulation. To improve detection efficiency, we employ an asymmetric optical duty cycle (e.g., 85%), thereby increasing the effective measurement time. Multiplying the sample signal by a square-wave reference signal (with levels ±1) and averaging effectively yields a weighted mean amplitude:

$$A_{ave}=\text{duty}\times A_{on}-(1\text{-duty})\times A_{off} \tag{S2}$$

where duty is the fraction of the modulation period during which the signal is “on,” and $A_{on}$ and $A_{off}$ are the signal amplitudes during the on and off states, respectively. The true difference between the on and off levels, $A_{true}=A_{on}-A_{off}$, can be recovered by correcting for the duty cycle of the modulation:

$$A_{true}=\frac{A_{ave}-(1-2\ \text{duty})\ A_{off}}{\text{duty}} \tag{S3}$$

In the common case where the off-level is negligible ($A_{off}=0$), this reduces to $A_{true}=A_{ave}/\text{duty}$.

For the optical excitation, the off-level is effectively zero ($A_{off}$=0), and the duty-cycle correction described above provides an efficient means to extract the true signal amplitude. A similar approach can be applied to the microwave modulation, which has $A_{off}\neq 0$, using a symmetric 50% duty cycle. However, complications arise when the optical and microwave modulation frequencies do not produce an integer ratio, making the duty-cycle correction less straightforward. In such cases, it is more practical to take advantage of post-processing: the “on” and “off” regions of the signal are identified directly from the reference waveforms. A time interval is considered “on” when both the optical and microwave modulation signals are positive, and “off” when the optical modulation is positive but the microwave modulation is negative, following the ±1 square-wave reference signals described above. The true signal amplitude is then obtained as the difference between the averaged on and off levels:

$$A_{true}=\langle A_{on}\rangle-\langle A_{off}\rangle \tag{S4}$$

To extract high-quality data, several experimental parameters must be carefully balanced. A primary consideration is the trade-off between sampling rate and averaging strength. Each point on the monitoring curve is allocated a fixed acquisition time, which limits the duration over which signal amplitudes can be averaged. This time window must be selected according to the kinetics of the monitored process; for the present measurements, a value of 4 s per point was chosen.

This allotted time can be distributed between lock-in demodulation averaging and trace accumulation. The number of modulation periods available for averaging is determined by the modulation frequency, which itself is physically constrained by the polarization and $T_1$relaxation times of the NV centers, typically on the order of ~1 ms. Lock-in demodulation requires multiple modulation periods within a trace to ensure sufficient averaging. A 0.4 s trace recorded at a 20 Hz optical modulation frequency contains eight optical periods, which is adequate for stable demodulation. The microwave modulation frequency was set six times higher, at 120 Hz, ensuring clear separation from the optical modulation.

At an optical duty cycle of 80%, five complete microwave periods occur during the light-on interval. At 120 Hz, each microwave cycle corresponds to approximately 8 ms, which is sufficiently long for the spin polarization response to approach its maximum amplitude within each cycle.

In principle, allocating 4 s per monitoring point allows ten successive 0.4 s traces to be accumulated. In practice, due to communication overhead and acquisition redundancies, approximately 8–9 traces can be acquired in rapid succession and subsequently averaged during post-processing on the PC.

The 500 MHz bandwidth oscilloscope enables acquisition of the 0.4 s doubly modulated trace at high sampling rates. We chose to record 1 million points per trace. However, approximately 1024 points per microwave period are already sufficient to accurately describe the highest modulation frequency. The oscilloscope's onboard averaging and decimation capabilities allow hardware-level averaging prior to data transfer, further improving signal quality and reducing communication overhead.

Lock-in demodulation is inherently sensitive, selectively suppressing noise components outside the reference frequency band. Nevertheless, slow parasitic drifts, such as light source intensity fluctuations, may still affect the measurement. By recording both the fluorescence modulation and the microwave modulation simultaneously, it is possible to calculate a contrast ratio, which further suppresses common-mode instabilities (that is, fluctuations that affect the optical excitation and microwave modulation signals proportionally) and enhances overall measurement stability.

## S5. Florescence and Microwave Amplitudes for High $O_2$ Sweep Range

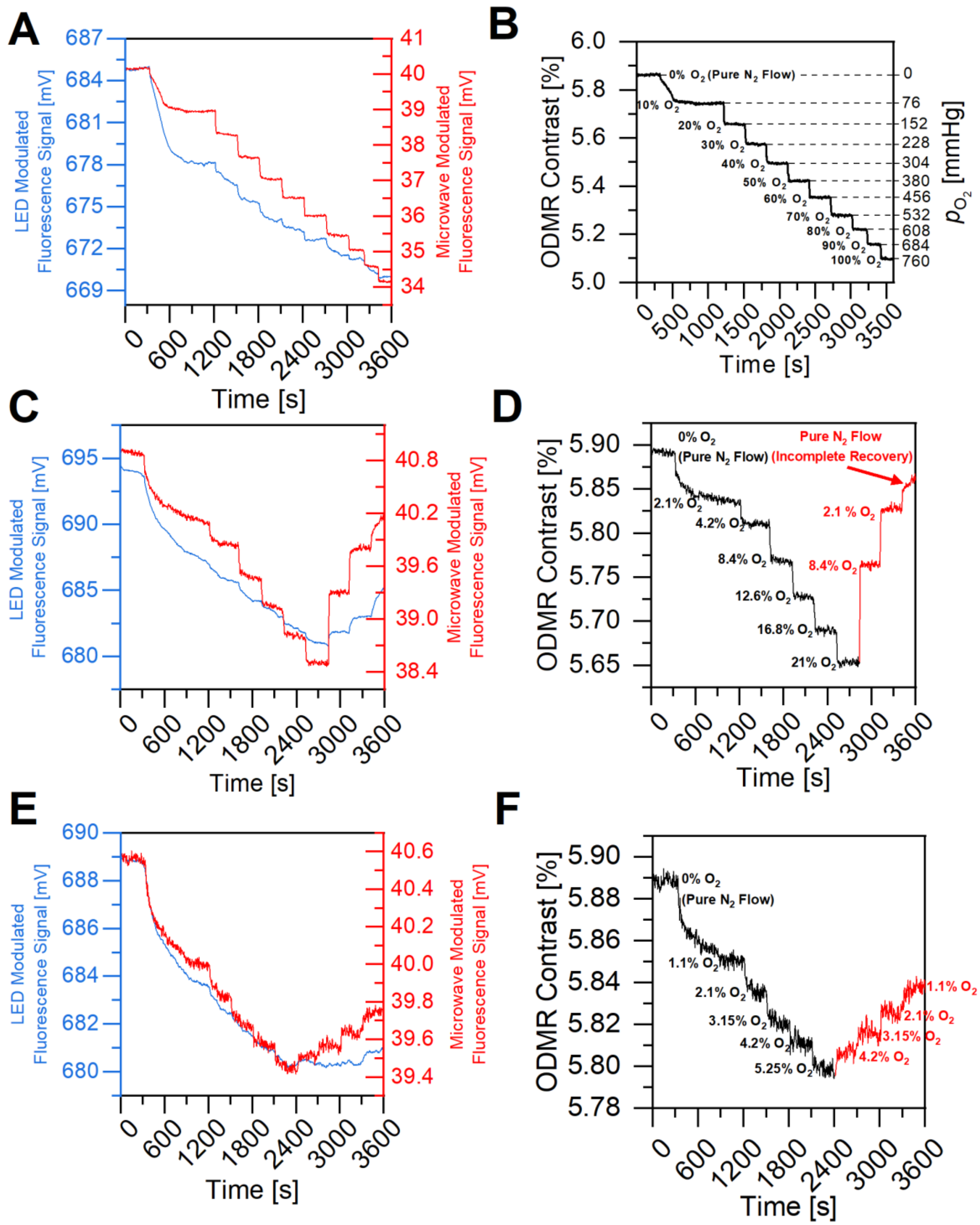

**Figure S5.** Recorded fluorescence and microwave signals for the high O2 sweep range (0-100%). The corresponding demodulated ratio is in Figure 2A in the main text. The blue PL Signal trace is used to calculated the Stern-Volmer plot in Figure 3 of the main text.

**S6. NV PL Spectra Under Either $N_2$ of $O_2$ Gas Flow**

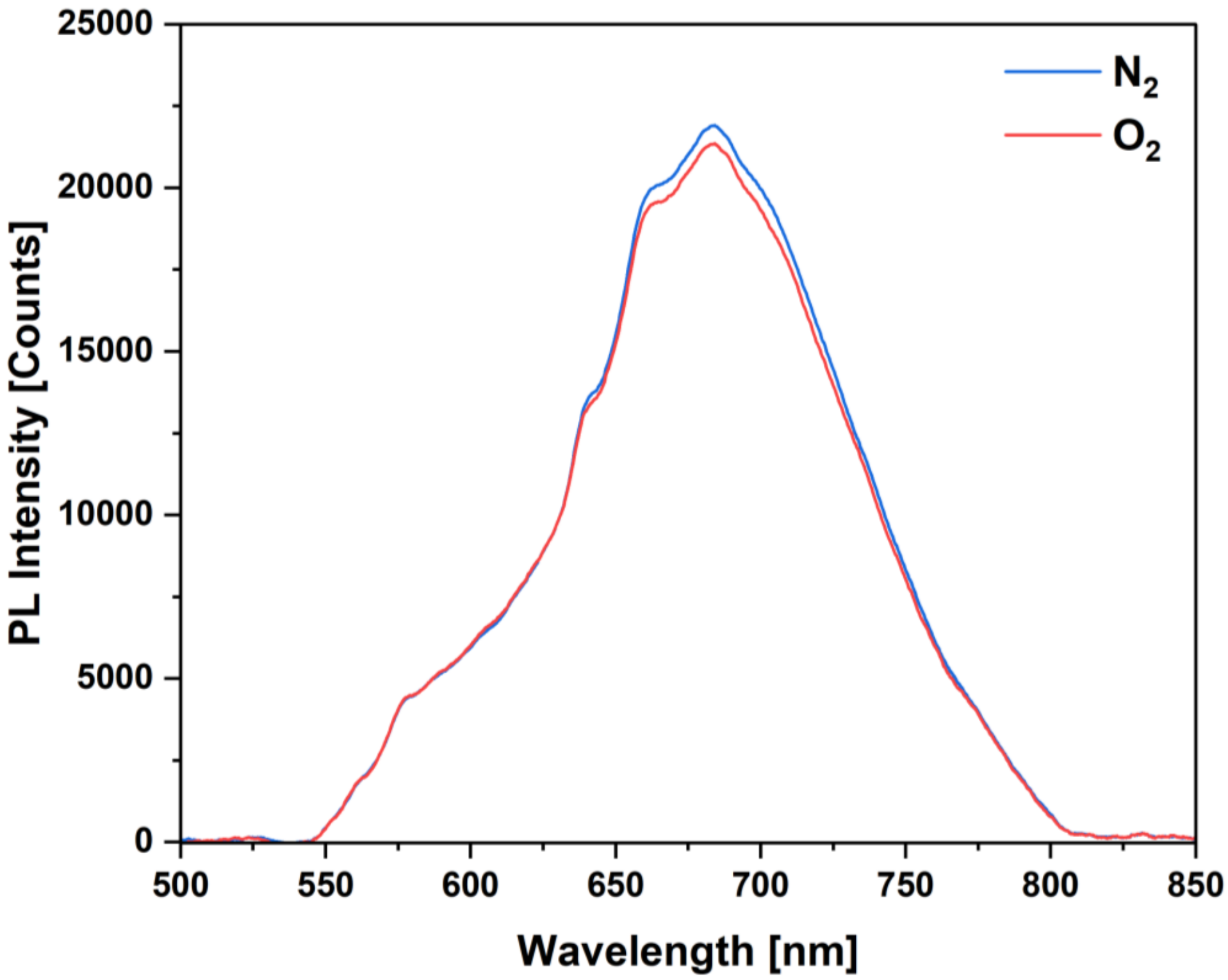

**Figure S6.** Photoluminescence spectra of 70 nm particles under either $N_2$ or $O_2$ flow.

## S7. Gas Mixture Experiment Details

**Table S1.** Experimental details of gas mixture experiments for high, intermediate, and low $O_2$ sweep ranges. Flow rates of gases are kept at 200 mL/min for all measurements.

| Fraction $O_2$ | Fraction $N_2$ | $p(O_2)$ [mmHg] | Time Range [s] | Corresponding Figure in Main Text |
|---|---|---|---|---|
| ***High $O_2$ Sweep Range*** | | | | |
| 0 | 1 | 0 | 0-300 | Figure 2A, B |
| 0.1 | 0.9 | 76 | 300-1200 | |
| 0.2 | 0.8 | 152 | 1200-1500 | |
| 0.3 | 0.7 | 228 | 1500-1800 | |
| 0.4 | 0.6 | 304 | 1800-2100 | |
| 0.5 | 0.5 | 380 | 2100-2400 | |
| 0.6 | 0.4 | 456 | 2400-2700 | |
| 0.7 | 0.3 | 532 | 2700-3000 | |
| 0.8 | 0.2 | 608 | 3000-3200 | |
| 0.9 | 0.1 | 684 | 3200-3400 | |
| 1 | 0 | 760 | 3400-3600 | |
| ***Intermediate $O_2$ Sweep Range*** | | | | |
| 0 | 1 | 0 | 0-300 | Figure 2C, D |
| 0.021 | 0.979 | 16 | 300-1200 | |
| 0.042 | 0.958 | 32 | 1200-1600 | |
| 0.084 | 0.916 | 64 | 1600-1900 | |
| 0.126 | 0.874 | 96 | 1900-2200 | |
| 0.168 | 0.832 | 128 | 2200-2500 | |
| 0.21 | 0.79 | 160 | 2500-2800 | |
| 0.084 | 0.916 | 64 | 2800-3100 | |
| 0.021 | 0.979 | 16 | 3100-3400 | |
| 0 | 1 | 0 | 3400-3600 | |
| ***Low $O_2$ Sweep Range*** | | | | |
| 0 | 1 | 0 | 0-300 | Figure 2E, F |
| 0.0105 | 0.9895 | 8 | 300-1200 | |
| 0.021 | 0.979 | 16 | 1200-1500 | |
| 0.0315 | 0.9685 | 24 | 1500-1800 | |
| 0.042 | 0.958 | 32 | 1800-2100 | |
| 0.0525 | 0.9475 | 40 | 2100-2400 | |
| 0.042 | 0.958 | 32 | 2400-2700 | |
| 0.0315 | 0.9685 | 24 | 2700-3000 | |
| 0.021 | 0.979 | 16 | 3000-3300 | |
| 0.0105 | 0.985 | 8 | 3300-3600 | |